\documentclass[preprint,superscriptaddress,preprintnumbers,prd]{revtex4}
\usepackage{latexsym}
\usepackage{amsthm}
\usepackage{amsmath}
\usepackage{graphicx}
\usepackage{amssymb}
\usepackage{multirow}
\usepackage{setspace}

\newcommand{\MeV}{\,\mathrm{MeV}}

\newcommand{\be}{\begin{equation}}
\newcommand{\ee}{\end{equation}}
\newcommand{\bea}{\begin{eqnarray}}
\newcommand{\eea}{\end{eqnarray}}
\newcommand{\f}[2]{\ensuremath{\frac{#1}{#2}}}
\newcommand{\bef}{\begin{figure}[htbp]\begin{center}}
\newcommand{\eef}{\end{center}\end{figure}}

\newcommand{\parf}[2]{\left(\frac{#1}{#2}\right)}
\newcommand{\mc}{\mathcal}
\newcommand{\gsim}{\lower.7ex\hbox{$\;\stackrel{\textstyle>}{\sim}\;$}}
\newcommand{\lsim}{\lower.7ex\hbox{$\;\stackrel{\textstyle<}{\sim}\;$}}

\begin{document}
\pagestyle{plain}
\preprint{SLAC-PUB-14197, FERMILAB-PUB-10-274-T}
\title{\Large Discovering New Light States at Neutrino Experiments}
\author{Rouven Essig}
\email{rouven@stanford.edu}
\affiliation{Theory Group, SLAC National Accelerator Laboratory, Menlo Park, CA 94025}
\author{Roni Harnik}
\email{roni@fnal.gov}
\affiliation{Theoretical Physics Department, Fermilab, Batavia, IL60510, USA}
\author{Jared Kaplan}
\email{jaredk@slac.stanford.edu}
\affiliation{Theory Group, SLAC National Accelerator Laboratory, Menlo Park, CA 94025}
\author{Natalia Toro}
\email{ntoro@stanford.edu}
\affiliation{Theory Group, Stanford University, Stanford, CA 94305}
\date{\today}
\begin{abstract}

Experiments designed to measure neutrino oscillations also provide major opportunities for discovering very weakly coupled states.  In order to produce neutrinos, experiments such as LSND collide thousands of Coulombs of protons into fixed targets, while MINOS and MiniBooNE also focus and then dump beams of muons.  The neutrino detectors beyond these beam dumps are therefore an excellent arena in which to look for long-lived pseudoscalars or for vector bosons that kinetically mix with the photon.  We show that these experiments have significant sensitivity beyond previous beam dumps, and are able to partially close the gap between laboratory experiments and supernovae constraints on pseudoscalars.  Future upgrades to the NuMI beamline and Project X will lead to even greater opportunities for discovery. 
We also discuss thin target experiments with muon beams, such as those available in COMPASS, and 
show that they constitute a powerful probe for leptophilic PNGBs. 

\end{abstract}

\maketitle

\tableofcontents

\section{Introduction}
In the last decade, several experiments have explored neutrino masses and mixings, but these 
high-luminosity laboratories are also sensitive to rare production of new metastable particles.
Neutrino beams such as the LAMPF Neutrino Source at Los Alamos and NuMI and BooNE at Fermilab 
are produced through two basic stages.  First, a high-intensity proton beam impinges on a target and 
produces a large number of pions (and other hadrons), which decay to
muons and neutrinos.  The muons are stopped in a thick layer of rock,
while the neutrinos travel unimpeded through the rock to the detector.
In fact, short-baseline neutrino detectors are situated behind 
the most intense proton \emph{and muon} beam-dumps to date.  Thus they
are ideally configured to search for long-lived particles produced by
rare proton-nucleus or muon-nucleus interactions.

Two classes of new physics scenarios naturally give rise to light,
feebly coupled particles of this type.  An approximate symmetry broken
at a high mass scale $F$ naturally gives rise to light pseudoscalars
--- pseudo-Nambu-Goldstone bosons (PNGBs, or ``generic axions'') ---
with couplings of order $m_X/F$ to Standard Model matter
$X$. Alternately, a new ``dark'' $U(1)$ gauge boson can naturally have
small kinetic mixing $\epsilon$ with the photon, giving rise to
suppressed interactions with all electrically charged matter
\cite{Holdom:1985ag}.  Either spin-1 or spin-0 bosons can be radiated
in energetic-particle interactions with matter, with a very small rate
proportional to the square of their weak coupling.  The luminosities
achieved in fixed-target experiments are such that thousands of these
particles could be produced, and they offer a new window into weakly
coupled sectors.  Once produced, the lightest particles in a hidden
sector could decay only through their weak couplings to Standard Model
particles, and so would be quite long-lived and weakly interacting.
While ordinary products of the collision are stopped in the shielding
upstream of a neutrino detector, exotics could penetrate the shielding
and decay within the detector, yielding a distinctive signal.

This note summarizes the several classes of exotic particles 
that naturally give rise to observable late-decay signals, their
experimental signatures and typical kinematics.  In any given model,
the production cross section and lifetimes of these exotica are both
determined by a single small coupling parameter and by the mass of the
produced particle, so in particular, we present specific estimates for
the sensitivity achievable with late-decay searches in MINOS/MINERvA,
MiniBooNE, and LSND.

PNGB's coupled to hadrons were searched for extensively
in both proton and electron beam-dump experiments in the 1980s, most
notably in CHARM \cite{Bergsma:1985qz} at CERN, E774 \cite{Bross:1989mp} at Fermilab, and 
the SLAC experiments 
E137 \cite{Bjorken:1988as} and E141 \cite{Riordan:1987aw}.  Many of these limits have recently been
re-interpreted  (\cite{Bjorken:2009mm,Batell:2009di,Schuster:2009au})
in the context of kinetically mixed gauge bosons and the associated
scalar bosons that give them mass through the Higgs mechanism.  
Kinetically mixed gauge bosons have been a subject of considerable recent interest \cite{Finkbeiner:2007kk, ArkaniHamed:2008qn, Pospelov:2008jd,
 Cholis:2008qq, Cholis:2008wq, Ruderman:2009tj, Dienes:1996zr,Cheung:2009qd,Katz:2009qq,Morrissey:2009ur,Cui:2009xq,Abel:2008ai,Ringwald:2008cu} and discussions of other collider, accelerator, and direct and indirect astrophysical probes for them can be found in e.g.~\cite{Batell:2009yf,Essig:2009nc,Bossi:2009uw,Yin:2009mc,Freytsis:2009bh,Baumgart:2009tn,Cheung:2009su,Abazov:2009hn,Batell:2009zp,Schuster:2009fc,Meade:2009mu,Yin:2009yt,Essig:2009jx,Galli:2009zc,Slatyer:2009yq,Essig:2010xa,Essig:2010em,Pospelov:2010cw}.
In
particular, \cite{Batell:2009di} discussed the sensitivity of
neutrino experiments to hadronic production in the case of kinetic
mixing and the potential importance of existing LSND
data as a constraint on these models.  Additional 
constraints on both classes of models from supernovaes, rare decays, 
and radiative corrections have also been extensively discussed
\cite{Pospelov:2008zw,Batell:2009jf, Freytsis:2009ct, Bjorken:2009mm, Andreas:2010ms, Goh:2008xz}.  

Our aim in this note is to present a more complete 
analysis of the sensitivity of past, present, and future neutrino experiments to
new weakly-coupled physics, with a particlar focus on PNGB models.  We hope
such a unified summary will facilitate new analyses of
neutrino-detector data to discover or constrain new weakly-coupled
particles.  
We also discuss the potential reach for experiments with muons beams that strike a fixed thin target.
In \S \ref{sec:otherconstraints}, we review constraints 
on generic pseudoscalars from other arenas, in particular supernova 
data, rare meson decays, and the anomalous muon magnetic moment.  
The range explored by neutrino experiments is 
complementary to all of these.  In \S \ref{sec:LSND}, we consider 
the implications of existing LSND analyses for both PNGB's and dark 
gauge bosons.  Due to the large number of protons dumped in LSND, 
we find that these analyses provide stronger constraints on 
PNGB's than the CHARM experiment.  
Our results for dark gauge bosons are consistent with 
\cite{Batell:2009di}, but we have tried to clarify the experimental sensitivity.  
In \S \ref{sec:modern}, we consider the sensitivity to PNGBs that could be achieved by analyses
using modern neutrino beamlines, such as BooNE and NuMI, and their
near detectors (the complementary analysis for kinetically mixed gauge bosons was presented in \cite{Batell:2009di}).  We consider both the standard production mode in
proton-nucleus collisions and the production of very forward PNGBs off
the stopping muons.  The second mode is enhanced by the magnetic
focusing of pions, so experiments using focused neutrino beams are
uniquely sensitive to purely leptophilic PNGBs that do not couple to
hadrons.  For ordinary PNGBs coupled to both quarks and leptons,
searches in MINOS/MINERvA and MiniBooNE would have sensitivity comparable to, 
or perhaps slightly better than, the CHARM beam-dump limit \cite{Bergsma:1985qz}. 
A future ``Project X'' could significantly extend this reach into new territory.
For leptophilic PNGBs, MINOS/MINERvA has slightly better sensitivity 
than the constraint from E137.  
In \S \ref{sec:thin}, we discuss muon fixed-target experiments using thin targets.  
Such an experiment could be possible at the COMPASS experiment at CERN. 
We find that a COMPASS-like setup can probe new territory in the parameter space of leptophilic PNGBs, closing a gap between muon $g-2$ limits and those from neutrino experiments.
We conclude in \S \ref{sec:conclusions}, and an appendix discusses the 
details of PNGB production off muon beams.

\subsection{Models of Weakly Coupled Light Exotics}

\subsubsection{Pseudo-Goldstone Bosons}

Light pseudoscalars can arise as pseudo-Goldstone bosons in a large variety of well-motivated theories, 
such as multiple higgs doublet models, theories with an R-axion \cite{Nelson:1993nf} (from spontaneous and explicit 
R-symmetry breaking in a supersymmetric theory), axion models \cite{Hagmann:2008zz}, the Next-to-Minimal Supersymmetric Standard model 
\cite{Dermisek:2005ar} (NMSSM), and recent dark matter models with a scalar portal to the dark sector \cite{Nomura:2008ru}.  
The most important point is that these particles are naturally light if there is an approximate shift symmetry, and 
that their interactions are proportional to the inverse of some symmetry breaking scale $F$.  Using 
fermion equations of motion, the derivative coupling of a PNGB $a$ to a fermion bilinear turns into the 
operator (in Weyl notation)
\be
L \supset \frac{m_\chi}{F} a \chi \chi ,
\ee
which is the coupling we assume for the leptons and/or quarks, $\chi$, of the Standard Model.  A 
phenomenologically interesting sub-class of PNGB models are those where the PNGB is leptophilic, 
i.e.~it couples preferentially (or only) to leptons; this scenario could 
arise if the lepton sector has its own higgs mechanism separate from that of the quark sector.  
Leptophilic PNGB are essentially unconstrained by searches for rare meson decays and proton fixed 
target experiments, so experiments that have muon beams, such as MINOS/MINERvA, MiniBooNE, and 
COMPASS, can easily be the most sensitive probes of these particles.  

\begin{figure*}[!t]
\begin{center}
\includegraphics[width=.45\textwidth]{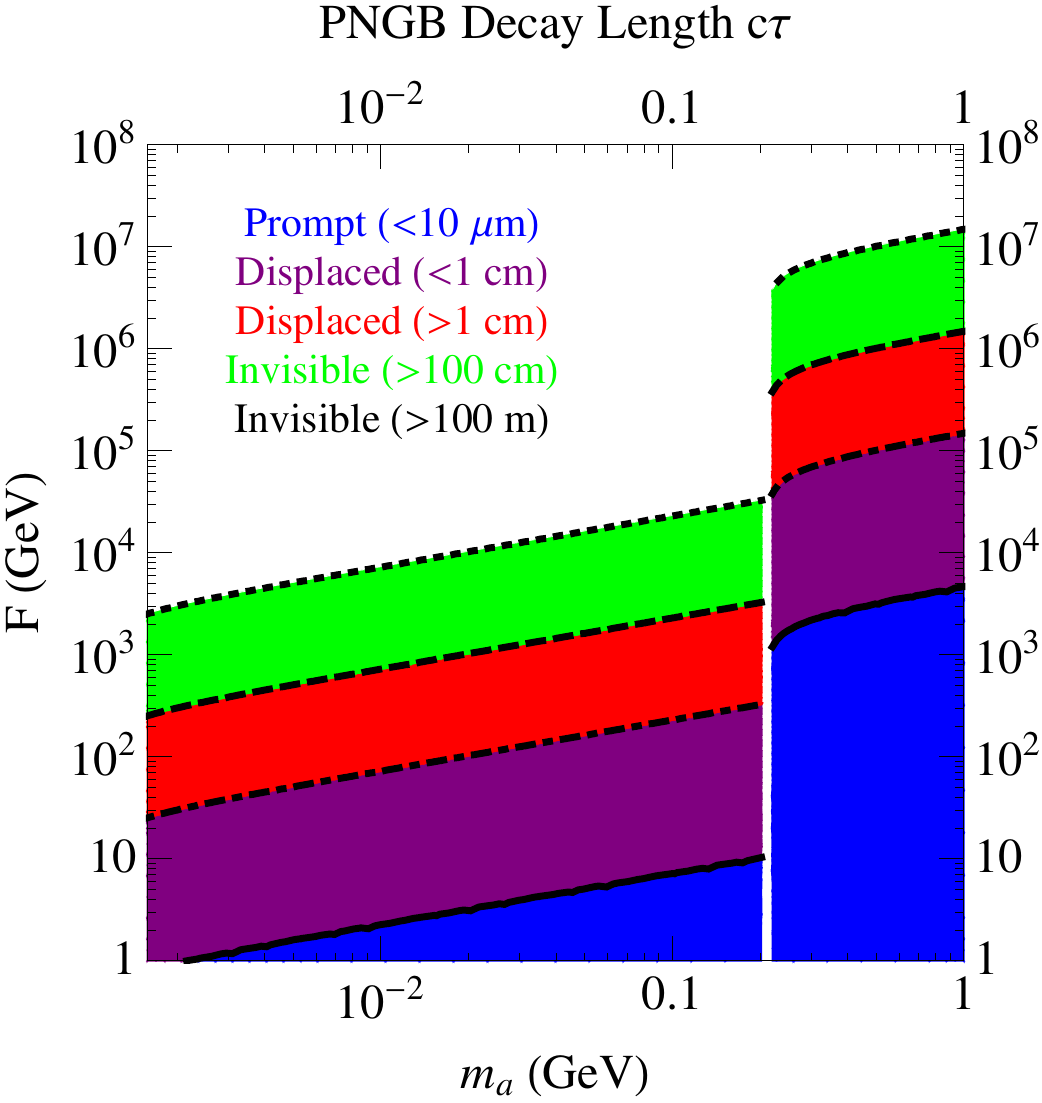}\;\;\;
\includegraphics[width=.47\textwidth]{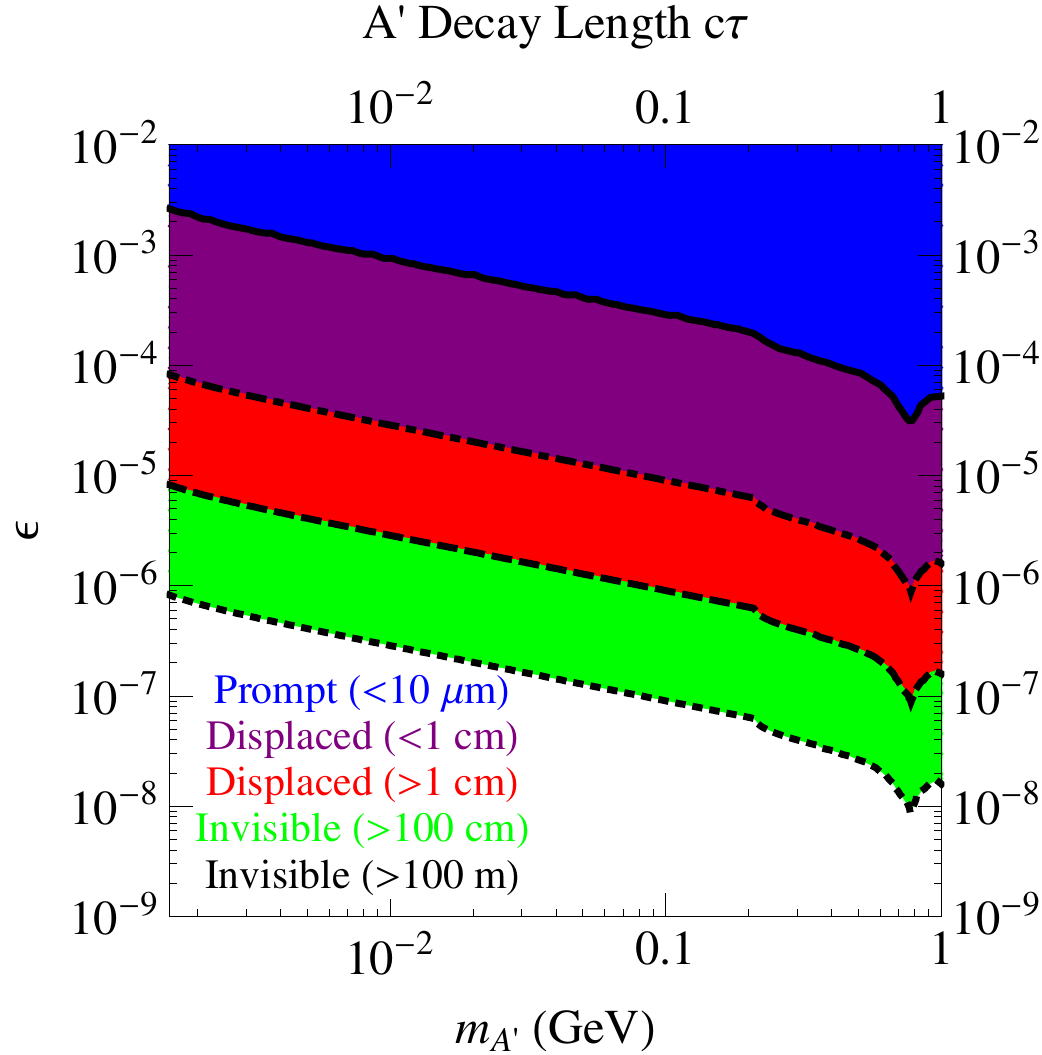}
\caption{{\bf Left:} The rest-frame lifetime of a pseudo-Nambu-Goldstone boson (PNGB) as a function of its 
mass $m_a$ and decay constant $F$. 
{\bf Right:} The lifetime of a dark photon $A'$ as a function of its mass $m_{A'}$ and 
$\epsilon$, the strength of its mixing with the Standard Model hypercharge gauge boson.
In both plots, the black lines correspond to different decay lengths ($c\tau$): 
$10~\mu m$ (solid), 1 cm (dot-dashed), 1 m (dashed), and 100 m (dotted).  
In the blue, purple, red, green, and white shaded regions the decays are prompt ($<10~\mu$m), 
displaced with $<1$ cm, displaced with $>1$ cm, ``invisible'' with $>100$ cm, or ``invisible'' 
with $> 100$m, respectively.
In fixed-target or beam dump experiments the particles 
typically get a large boost that increases their decay length by $E_{\rm beam}/$mass. 
The feature in the left plot at 2$m_\mu$ occurs since PNGB's coupling to a Standard Model 
particle is proportional to that particle's mass, and at this point decays to two muons are allowed.  
The dip in the right plot near 0.7 GeV is due to the $\rho$-resonance.  
The lifetime for both the PNGB and the $A'$ is calculated assuming decays directly into 
Standard Model particles.
}
\label{fig:Lifetime}
\end{center}
\end{figure*}

The coupling of a PNGB with mass $m_a$ to a fermion with mass $m_\ell$ induces a PGNB partial 
width
\be\label{eq:PNGBwidth}
\Gamma_\ell = \f{m_a}{8\pi} \parf{m_\ell}{F}^2 \sqrt{1 - (4 m_\ell^2/m_a^2)}
\ee
and the total width is well approximated by $\Gamma_e + \Gamma_\mu$ for 
$m_a \lesssim 400$ MeV (for larger masses, hadronic decays can also become important but we 
use the leptonic widths for masses up to 1 GeV).  Thus, for example, proper lifetimes of 1 mm are 
obtained with $F \approx 70$ GeV ($m_a = 100$ MeV) or $F \approx 20$ TeV ($m_a = 300$ MeV).
Fig.~\ref{fig:Lifetime} (left plot) shows the decay length ($c\tau$) of a PNGB as a function of 
its mass $m_a$ and decay constant $F$.  
Note that the decay length is very different above and below the muon threshold, due to the much
stronger coupling to muons as compared to electrons.  We see that for $F\lesssim 10^2$ GeV, they 
decay promptly 
and colliders should be able to set the best constraints.  For larger $F$, collider searches 
that look for displaced vertices or missing energy can still set limits, but searches in 
beam dump experiments (with a large shield) become relevant.  

We ignore decays of PNGBs to two photons, since this is always subdominant in the mass range we
consider in this paper. 

\subsubsection{Kinetically Mixed Gauge Bosons}

Another class of light particles that has received significant interest
in the last few years is a light, weakly coupled ``sequestered
sector'' lying alongside the Standard Model.  New states at the
MeV--GeV scale are not in conflict with data, because the gauge and
global symmetries of the Standard Model greatly restrict the couplings
of ordinary matter to these states.  It is, however, quite natural for
the few allowed interactions to be suppressed by loop factors.  

In particular, we consider a new MeV--GeV scale Abelian 
gauge boson $A'$ coupled to electrically charged Standard Model particles $\psi$
\be\label{eq:KM} 
\delta {\cal L}  = \epsilon e A'_\mu \bar \psi\gamma^\mu \psi. 
\ee 
Such a coupling can generically originate from the kinetic mixing between the field strengths of 
the Standard Model hypercharge and a hidden sector $U(1)$~\cite{Holdom:1985ag},
\be\label{kmix} \delta {\cal L} = \frac{\epsilon_Y}{2} F'_{\mu\nu}
F^{\mu\nu}_Y, \ee where
$F'_{\mu\nu}=\partial_{\mu}A'_{\nu}-\partial_{\nu}A'_{\mu}$ is the
field strength of the $A'$ gauge boson, and similarly $F^{\mu\nu}_Y$
is the hypercharge field strength.  
This mixing, assuming the hidden $U(1)$ is broken so that the $A'$ is massive, 
is equivalent in low-energy interactions to assigning a
hidden charge $\epsilon e q_i$ to Standard Model particles of electromagnetic
charge $q_i$, where $\epsilon = \epsilon_Y/(\cos\theta_W)$ and
$\theta_W$ is the Weinberg mixing angle (Eq.~(\ref{eq:KM}) if $q_\psi = 1$) .  
Such a mixing can be generated in many ways, e.g.~through loops of new heavy 
particles that couple to both the $A'$ and Standard
Model hypercharge.  We refer the reader to e.g.~\cite{Baumgart:2009tn} for 
more detailed discussions.

In Fig.~\ref{fig:Lifetime} (right), we show the decay length of a vector boson $A'$ as a function of 
its mass $m_{A'}$ and the parameter $\epsilon$ which sets the strength of its kinetic mixing with 
hypercharge.  
Assuming the $A'$ decays into Standard Model particles rather than exotics, its lifetime is 
\bea
\gamma c\tau \simeq \f{3}{N_{\rm eff} m_{A'} \alpha \epsilon^2} 
\simeq \f{0.8\mbox{cm}}{N_{\rm eff}} \left ( \f{E_0}{10 \mbox{GeV}} \right ) \!\!
\left (\f{10^{-4}}{\epsilon} \right )^2 \!\!
\left ( \f{100\, \mbox{MeV}}{m_{A'}} \right )^2,
\label{gammaCTau}
\eea
where we have neglected phase-space corrections, and 
$N_{\rm eff}$ counts the number of available decay products.  
If the $A'$ mixes kinetically with the photon, then $N_{\rm eff} = 1$ for $m_{A'}  < 2
m_{\mu}$ when only $A'\to e^+e^-$ decays are possible, and $2+R(m_{A'})$ 
for $m_{A'}\ge2 m_{\mu}$, where 
$R=\f{\sigma(e^+e^- \rightarrow \mbox{ hadrons};\, E=m_{A'})}{\sigma(e^+e^-  \rightarrow \mu^+\mu^-;\, E=m_{A'})}$ 
\cite{Amsler:2008zzb}.  


\begin{table}[!t]
\begin{tabular}{l@{\qquad}p{1.4in}@{\qquad}p{1.4in}@{\qquad}p{1.6in}}
\hline
Particle & Final state & Production mode & Momentum spectrum \\
\hline
PNGB & 
$\begin{cases}
2\mu & (m > 2 m_\mu) \\
2e,\, 2\gamma & (m < 2 m_\mu) \end{cases}$
& Proton-Nucleus ($a/\pi^0$-mixing)& Same as $\pi^0$ spectrum \vspace{24pt}\\ 
'' & " & Muon $a$-strahlung & muon spectrum (avg. over material) \\
\hline
Dark gauge boson & 
$2e$, $2\mu$, hadronic modes 
& Proton-Nucleus \quad ($\pi^0$ decay) & $1/2$ of typical final-state $\pi^0$ momentum \vspace{12pt}
\\ 
" & " & Muon-Nucleus \quad ($A'$-sstrahlung) & muon spectrum (avg. over material) \\
\hline
\end{tabular}
\caption{Summary of Signal Properties for Light Exotics considered in this paper.  The PNGB decay 
to two photons is never important for the mass range we consider, and can be ignored. More 
complicated signals are possible if the PNGB or $A'$ can decay to other hidden sector 
particles before decaying to Standard Model particles, but we will not consider this possibility.}
\label{tab:signal}
\end{table}

\vskip 5mm 
We summarize the possible production and decay mechanisms for both PNGBs and $A'$'s in 
Table \ref{tab:signal}.
In this paper, we only consider the case for which the PNGB or $A'$ decays directly to Standard
Model matter.  More complicated signals are possible if they can decay to other hidden sector 
particles before decaying to Standard Model particles.  

\section{Constraints on PNGBs from Rare Meson Decays and Supernovas}\label{sec:otherconstraints}

\subsection{Limits from Kaons and B-Meson Decays}\label{sec:mesons}

\begin{table}
\begin{tabular}{|l|c|c|c|}
\hline Decay mode & B.R.~limit & Mass range & Decay region \\ \hline 
$B^+ \rightarrow K^+ + inv.$ \cite{Belle:2007zk} & $3\times10^{-6}$ & $0 <m_a < 2 m_\mu^{(*)}$ & $>1.65$m \\ 
$K^+ \rightarrow\pi^+ + inv.$ \cite{Adler:2004hp}  & $0.5-10 \times 10^{-10}$ & $0 <m_a < 2 m_\mu^{(*)}$ & $>1.3$m \\
$K^+ \rightarrow \pi^+ + X\;X\rightarrow e^+e^-$ \cite{Yamazaki:1984vg}& $1.5-5\times10^{-6}$ & $10 \MeV < m_a < 120 \MeV$ & $<5$cm$^{(\dagger)}$ \\
$B^+ \rightarrow K^+ e^+ e^-$ \cite{Aubert:2005cf} & $6.7\times 10^{-7}$ & $30 \MeV < m_a < 2 m_\mu^{(*)}$ & $<100\mu$m$^{(\dagger)}$ \\
$d\Gamma/dm^2(B^+ \rightarrow K^+ \ell^+ \ell^-)$ \cite{Belle:2008sk} & $1.2\times 10^{-7}$ & $140 < m_a < 1440$ MeV & $<.5$cm \\
\hline
\end{tabular}
\caption{\label{tab:meson}
Summary of constraints on PNGBs from various experiments (B.R.~= Branching Ratio).  For decay of meson $X$, a characteristic transverse energy $m_X/2$ is assumed in computing the fraction of decays either in or outside a cylinder whose radius is given in the ``decay region'' column.
\\$^{(*)}$: Searches extend beyond $2 m_\mu$ but only imply relevant limits on PNGB models below $2 m_\mu$.\\
$^{(\dagger)}$: Length scales are guesses only, but overall exclusion is insensitive to cutoff because it overlaps with reach of invisible searches.}
\end{table}

In this section and in Table \ref{tab:meson}, we briefly summarize
constraints from meson decays on PNGBs with sub-GeV masses.  These
constraints are controlled by two factors: the partial width for the
rare meson decays into a PNGB, and the fraction of PNGB's that decay
promptly enough to be included in the data samples (or, in the case of
invisible-decay searches, the fraction that decay outside the
detector).  The combination of these searches is sensitive to PNGBs
with $F\lesssim 10-100$ TeV over a wide range of masses.  We call
these limits ``exclusions'' below in the interest of brevity, but it
should be emphasized that several of the results have been
re-interpreted by non-experts (the authors) in a context very
different from the original experimental design, from which
significant inaccuracies in the boundaries could have resulted.  We
focus here on the most constraining searches; a more exhaustive set of
limits are considered in \cite{Andreas:2010ms}, but our treatment of
the decay regions and the PNGB lifetime differ from theirs.

Following \cite{Freytsis:2009ct}, we consider two-higgs-doublet models
as a generic framework for a new PNGB $a$ that couples to the Standard Model
(whereas \cite{Andreas:2010ms} assumes NMSSM couplings).  In this
framework, rare meson decays to a lighter meson and $a$ are
mediated primarily by a top-quark loop, which receives contributions at all
scales up to the highest energy at which the PNGB couples to the top quark (typically
either the $F$ scale or the electroweak symmetry-breaking scale). Therefore, 
the meson decay rates depend on the detailed UV dynamics through which the $a$
particle couples to Standard Model fields, as discussed in
\cite{Freytsis:2009ct}.  We adopt the approximate formula
\be
\Gamma(B^+\rightarrow K^+ a) \approx \frac{G_F^3 |V_{tb}^* V_{ts}|^2 }{\sqrt{2} 2^{12} \pi^5} m_t^4 m_B^3 \parf{2 v^2}{F^2} (kinematic) [f_0(m_a)^2], 
\ee
and likewise for $K^+\rightarrow \pi^+ a$,  but with the product of CKM 
matrices $V_{tb}^*V_{ts}$ instead given by  $V_{ts}^*V_{td}$.  
Here $G_F$ is the Fermi coupling constant,
$v=174$ GeV is the electroweak Higgs vacuum expectation value, $m_t$ is the top mass, and
$m_B$ the $B$-meson mass.  This formula is obtained from (A.1) of
\cite{Freytsis:2009ct} by assuming $\beta = 45^\circ$ in the
two-higgs-doublet model, replacing $\sin \theta \rightarrow v/(2 F)$,
and setting the model-dependent combination $|X_1+X_2|=1$.  This
combination depends on the physical Higgs mass but is typically in the
range 1--10 so that our choice is conservative (although note that there are special
choices of the Higgs mass where the sum of $X$'s crosses zero, in
which case limits from meson decay are weaker).  The kinematic factor
is typically in the range $0.5$-$1$ and is defined in (A.3) of
\cite{Freytsis:2009ct}, while the form factor $f_0 \approx 0.33$
\cite{Ball:2004ye}.  The formula above with our parameter choices 
yields branching fractions about a factor of 10 smaller than those in
\cite{Andreas:2010ms}.

The second factor that comes into play is the decay length of the
PNGB.  For visible meson decay searches, the PNGB must decay within a
small distance (typically of order mm-cm) from the meson decay from
which it originates, so that all of the tracks produced by the meson
reconstruct a single vertex.  Likewise, invisible-decay searches
require that the PNGB decay \emph{outside} the detector volume, with a
typical size of about a meter.  We have attempted to take these
effects into account more precisely than \cite{Andreas:2010ms}, which
uses a uniform length scale of 10 m for all limits.  


The PNGB partial width to $\ell^+\ell^-$ is given in Eq.~(\ref{eq:PNGBwidth}) and the lifetime 
is shown in Fig.~\ref{fig:Lifetime}.  
Naively, one would expect that PNGB's with characteristic lifetimes between 1 mm
and 1 m might be poorly constrained by both visible-decay and
invisible-decay searches.  However, the fraction of events in which
the PNGB decays within an atypically short distance from the meson vertex can
still set significant constraints; moreover, the limits from
experiments with different energy scales ($m_K$ and $m_B$) overlap to
fill this intermediate-lifetime region.

\paragraph{Invisible decay searches} for $B^+\rightarrow K^+ \nu \bar\nu$ or $K^+ \rightarrow \pi^+ X$, where $X$ decays invisibly, peak in a specified mass range.  Below the muon threshold, these are sensitive to the highest $F$'s probed in accelerator experiments. 
\begin{itemize} 
\item $B^+\rightarrow K^+ + inv.$: A search at Belle
  \cite{Belle:2007zk} set an upper limit of $1.4\times 10^{-5}$ on the
  rate of $B^+ \rightarrow K^+ \nu\bar\nu$.  This search imposes an
  upper limit on the energy deposited in the electromagnetic calorimeter
  (ECL), which extends to 1.65 m from the beamline.  Therefore, for this limit we consider only $a$ decays outside this radius, assuming a typical transverse momentum of $m_B/m_a$.  A tighter limit could likely be set by searching for a narrow invisible resonance in the $B^+$ decays.
\item $K^+\rightarrow \pi^+ + inv.$: The search for $K^+ \rightarrow \pi^+ \nu \bar\nu$ at Brookhaven E787 \cite{Adler:2004hp} also set an explicit limit on the decay $K^+ \rightarrow \pi^+ X$ where $X$ is invisible.  The branching fraction for this mode must be below  about $5\times 10^{-11}$ for $X$ masses below about 100 MeV, and about $10^{-9}$ for $X$ masses between 150 and 250 MeV.  No limit is set between 100 to 150 MeV, where several $K^+ \rightarrow \pi^+ \nu \bar\nu$ candidates were seen.  The transverse size of the detector is roughly 1.3 m, and the kaons are produced at rest so that $m_K/2$ is a good approximation to the energy of the outgoing PNGB.
\end{itemize}
Of these two, the $K^+$ decay limit extends to higher $F$ but the $B^+$ decay search is able to probe lower $F$ because the PNGBs produced in $B$ decays are more boosted and therefore longer-lived.  There are additional constraints from CLEO and BaBar searches for $\Upsilon(1s)$ and $\Upsilon(3s)$ decays to $\gamma a$ \cite{Balest:1994ch,:2008hs,Sanchez:2010bm}, but this region is largely contained in the two identified above.

\begin{figure*}[!t]
\begin{center}
\includegraphics[width=.48\textwidth]{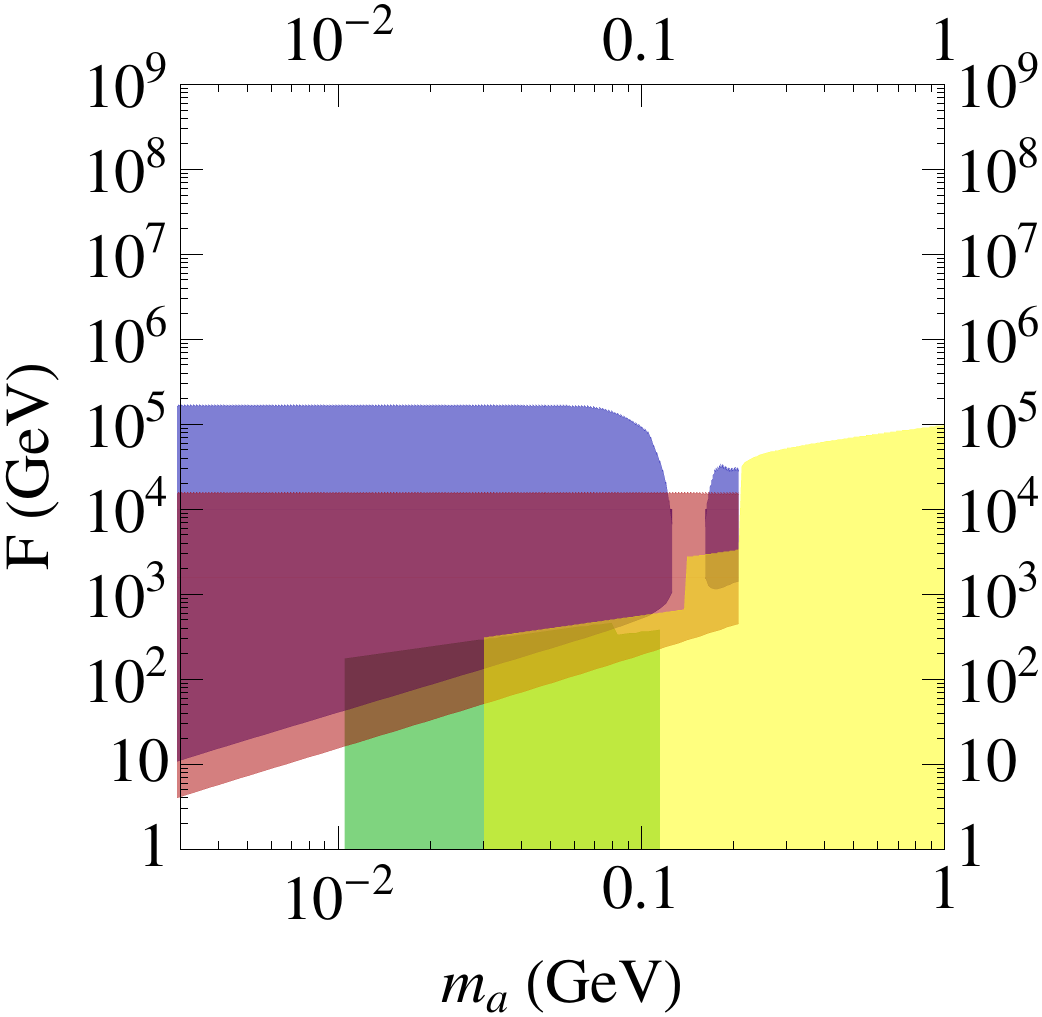} \;\; 
\includegraphics[width=.48\textwidth]{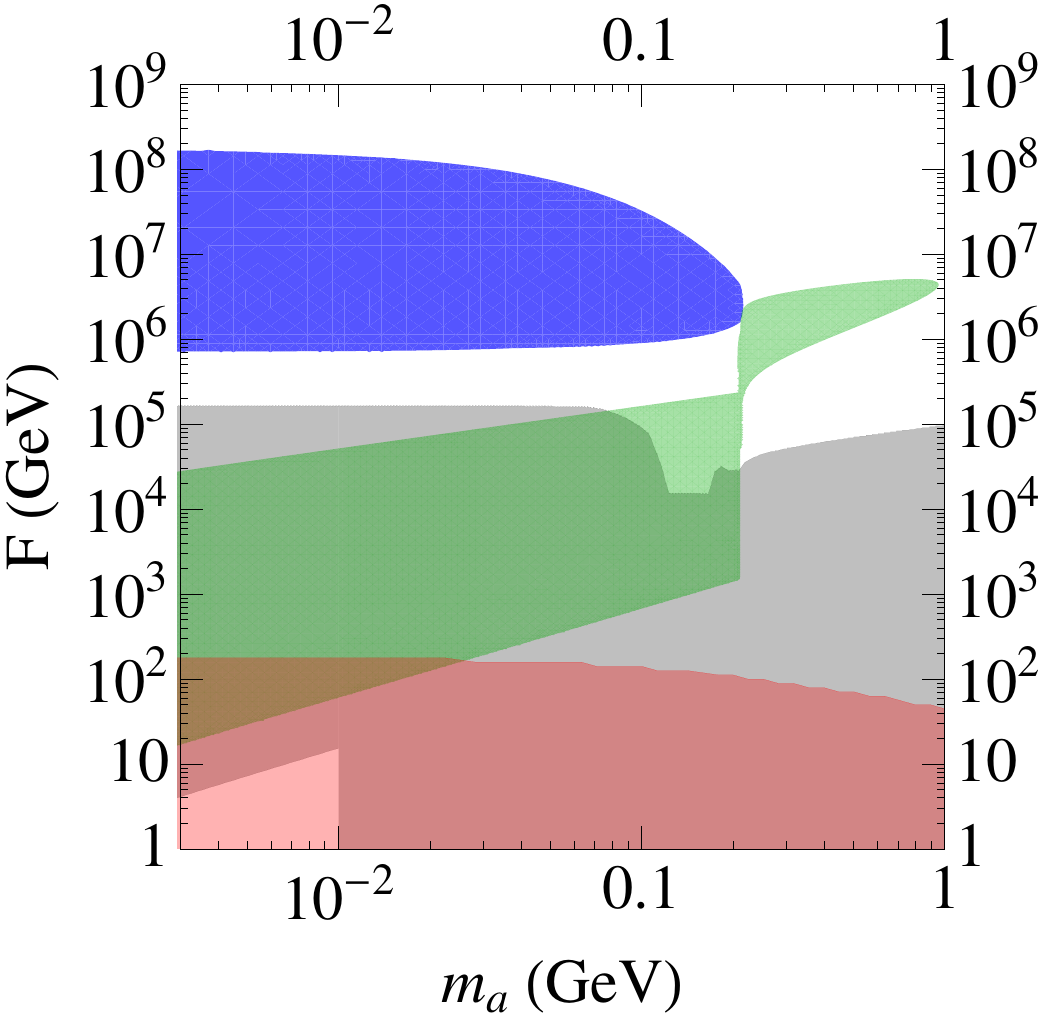}
\caption{{\bf Left:} Constraints on pseudo-Nambu-Goldstone bosons 
as a function of their decay constants $F$ and their mass $m_a$ 
from various meson decays: 
$K^+\to {\rm anything}+e^+e^-$ (green), $K^+\to \pi^++$~invisible (blue), $B^+\to K^+ \ell^+\ell^-$ (yellow) 
($\ell = e, \mu$), and $B^+\to K^+ + $~invisible (red).  
Constraints from $\Upsilon(1S)$ or $\Upsilon(3S) \to \gamma a \to \gamma +$ invisible 
and $K^+\to \pi^+ \ell^+\ell^-$ 
decays are weaker than those from 
$B^+\to K^+ + $~invisible and $B^+\to K^+ \ell^+\ell^-$, respectively,  
and thus not shown.  Details are in \S \ref{sec:mesons}.
{\bf Right:}
Gray shaded background region is the combined exclusion region from meson decays 
in the left figure. 
In the green exclusion region, the proton beam dump experiment CHARM at CERN 
would have seen at least five events (this exclusion region agrees roughly with that in 
\cite{Bergsma:1985qz}) -- see \S \ref{sec:modern}.  
Here the PNGB is produced directly in the proton dump by a small mixing with the pion.  
For $m_a < 2 m_\mu$, the PNGB decays to an electron pair, while in the ``bubble''  for 
$m_a>2 m_\mu$ the PNGB decays predominantly to a muon pair. 
The blue region is the limit from the supernova SN 1987a (see \S \ref{sec:SN}).  
The light red region is the constraint from the muon anomalous magnetic moment and 
fills the gap for low $m_a$ and $F$ left by the meson constraints (see \S \ref{sec:amu}).  
The region excluded by the Fermilab E137 dump lies mostly within the CHARM 
excluded region and is not shown (it is instead shown in Fig. \ref{fig:comboLepto}).
}
\label{fig:constraints}
\end{center}
\end{figure*}

\paragraph{Visible decays}
Again, we focus here only on the most powerful visible-decay searches:
\begin{itemize}
\item $K^+ \rightarrow \pi^+ X,\; X\rightarrow e^+e^-$: This resonant
  decay mode was searched for in the $K_{\mu 2}$ experiment at KEK
  \cite{Yamazaki:1984vg}, which excluded kaon branching fractions
to below about $1.5\times 10^{-6}$ for PNGB masses between 10 and 80
  MeV, and below about $5\times 10^{-6}$ for PNGB masses from 80 to
  120 MeV.  The experiment also set a limit above 140 MeV, but it is
  not as constraining as the $B^+$-decay limits discussed below, 
  so we do not include the constraint in this region for this mode.
  $K_{\mu 2}$ used stopped kaons, so we assume an initial energy of
  $m_K/2$.  No vertex requirement is explicitly mentioned in the
  analysis, but common vertex requirements of order a few cm are
  frequently imposed in spectrometer analyses typical in these
  experiments; in any case, at much larger distances, the mass resolution
  would likely be degraded.  We have used an \emph{ad hoc} but
  conservative estimate of 1 cm to produce the limit in Fig.~\ref{fig:constraints} 
  but even if the vertex requirement is much tighter (as
  tight as $0.5$ mm) the excluded region would overlap that of the
  $B^+ \rightarrow K^+ + inv.$ search.
\item $B^+ \rightarrow K^+ \ell^+\ell^-$: BaBar \cite{Aubert:2005cf}
  and Belle \cite{Ishikawa:2003cp, Belle:2008sk} have both measured the
  rate of the rare decay $B^+ \rightarrow K^+ \ell^+\ell^-$ for
  $\ell=e,\, \mu$.  The observed branching fractions, (3--6)$\times
  10^{-6}$, are consistent with Standard Model predictions, and can be
  translated into rough (conservative) limits on $B^+ \rightarrow K^+
  a$ by requiring that this exotic decay not exceed the total measured
  rate (we focus on the decays to $K^\pm$ rather than $K^*$ because
  the observed rates are lower).  The BaBar measurement includes
  lower-mass electron pairs, down to 30 MeV (compared to 140 MeV at Belle).
  The most recent Belle analysis \cite{Belle:2008sk} bins events by
  invariant mass, so that we can obtain a tighter limit ($\approx
  10^{-7}$) on $B^+\rightarrow K^+ a$ in the region of interest,
  140-1440 MeV.  In all cases we take the limit to be the central
  value plus $2\sigma$. The BaBar measurement required that the
  $\ell^+\ell^-$ pair originate from the same vertex as the $K^+$.  To
  set a conservative limit we require that the $a$ decay within
  $100\mu$m, the scale of BaBar's vertex resolution. For the Belle
  analysis we require only that the PNGB decay within 0.5 cm, which
  would pass the requirement of \cite{Ishikawa:2003cp}.
\item Similar but slightly weaker limits are obtained from measurements 
of $K\rightarrow \pi \ell^+\ell^-$, e.g.~\cite{Appel:1999yq,Goudzovski:2009mg,Marinova:2010ay}.  
We refer the reader to the original results and to  \cite{Andreas:2010ms} for details.
\end{itemize}
It is worth emphasizing that the crude limits we have obtained are far
weaker than the tightest limits that \emph{could} be obtained by a
directed analysis of BaBar or Belle data.  Firstly, much tighter
limits could be obtained by binning the 20--100 events in each sample
more finely, and accounting for the detection efficiency as a function
of mass.  Further improvement could be obtained by including more
displaced $a$ decays, at the edge of the inner tracker (3 cm) or even
beyond.  \emph{This direction is particularly worthy of exploration for the
$B\rightarrow K \mu^+\mu^-$ mode, for which there are no complementary
searches near the high-$F$ boundary of the Belle-excluded region, and
for which the muon system gives an additional handle for studying
highly displaced decays.}

\subsection{Limits from Supernova SN 1987a}\label{sec:SN}

For completeness we also include the constraints on PNGBs from SN 1987a \cite{Raffelt:1990yz}.  We adapt the analysis of \cite{Burrows:1988ah} to our setup, and obtain limits based on the assumption that PNGBs must not be the dominant mechanism of energy loss from the supernova.  The temperature of the supernova core is conservatively estimated at $T \sim 30$ MeV \cite{Burrows:1988ah}, so PNGBs with mass significantly greater than these energies cannot be produced by the supernova, and are therefore unconstrained.  The flux of PNGBs from the core of the supernova is approximately \cite{Burrows:1988ah}
\be
\frac{d N_a}{dE_a} \sim 10^{71} \left( \frac{1 \ \mathrm{GeV}}{F} \right)^2 e^{-E_a/T} 
\frac{1}{\mathrm{GeV}}\;.
\ee
However, if these PNGBs decay or are re-absorbed, then they will not escape from the supernova and so they will not carry away any energy.  Scattering and re-absorption dominate over PNGB decay for the relevant region of parameter space, giving a mean free path of
\be
\lambda_{\rm mfp} \sim 10 \ \mathrm{m}  \left( \frac{F}{10^6 \ \mathrm{GeV}} \right)^2
\ee
for PNGBs with $F  < 10^8$ GeV.  We see that for $F$ significantly smaller than $10^6$ GeV, the PNGB mean free path is much less than the estimated core size of $10$ km, so for these smaller values of $F$, SN 1987a does not constrain the PNGB.  The exclusion contours we have plotted in 
Fig.~\ref{fig:constraints} (right) correspond to requiring that the PNGBs carry away less energy than neutrinos, meaning that the total integrated PNGB emission must be less than about $10^{53}$ PNGBs, each with energy of order $T$.

\subsection{Limits from the anomalous muon magnetic moment}\label{sec:amu}

PNGB's contribute to the anomalous magnetic moment of the muon, $a_\mu$, at the loop level.  
For the mass range of interest in this paper ($\lesssim 1$ GeV), only the one-loop 
contribution is important, and it is given by (see e.g.~\cite{Domingo:2008bb,Andreas:2010ms})
\be
a_\mu^a = -\f{1}{8\pi^2} \, \f{m_\mu^2}{F^2} \; \int_0^1\,dx\, \f{x^3}{x^2+(1-x)\f{m_a^2}{m_\mu^2}}.
\ee
While the experimental measurement of $a_\mu$ is rather 
precise \cite{Bennett:2006fi}, the Standard Model prediction involves a 
hadronic contribution that must be estimated from experiments, which do not all agree.  
Using data from $e^+e^-$ annihilation to hadrons, the theoretical value of $a_\mu$ 
is smaller than the measured value by ($316\pm 79$)$\times 10^{-11}$ \cite{Teubner:2010ah}, 
which is a $4\sigma$ discrepancy.  
However, estimates from $\tau$'s give a smaller disagreement, with 
\cite{Davier:2009ag} finding a difference of ($157\pm 82$)$\times 10^{-11}$, which is a 
1.9$\sigma$ discrepancy. 

Since the contribution from PNGB's is negative, a very conservative limit is obtained 
by using the $5\sigma$ lower bound in \cite{Davier:2009ag}, i.e.
\bea
a_\mu^a & \ge & (157-5\times 82) \times 10^{-11}, ~\textrm{i.e.} \nonumber \\
a_\mu^a & \ge & -253 \times 10^{-11}.
\eea
This constraint is included in Fig.~\ref{fig:constraints} (right) and \ref{fig:comboLepto}.

\section{Constraints from LSND on PNGBs and Dark Gauge Bosons}\label{sec:LSND}

The Liquid Scintillator Neutrino Detector (LSND) experiment ran at the
Los Alamos Neutron Scattering Center (LANSCE) in the 1990's
\cite{Athanassopoulos:1996ds}, and dumped $\mc{O}(10^{23})$ 800 MeV
protons on a predominantly water-copper target.  This produced $\sim
10^{22}$ pions, a very large number that allows LSND to be sensitive
in principle to very weakly coupled PNGBs or gauge bosons.  A detector
of length 8.3 m and a diameter of 5.7 m was located 29.7 m away from
the target, $12^\circ$ off-axis~\footnote{The KARMEN experiment~\cite{Armbruster:2002mp} had similar features and also dumped a large number of protons, above 9000 $C$. However the KARMEN detector was $90^\circ$ off axis and is thus not useful for our purposes.}, and filled with dilute liquid
scintillator (there was no open decay region in front of the
detector).  Although the LSND collaboration did not search for signals
that could originate from decays of long-lived exotics, approximate
limits can be extracted from two published LSND analyses.

We begin by reviewing the production of PNGB's in proton-nucleus
collisions (which will re-appear in our discussion of more recent
neutrino experiments in \S \ref{sec:modern}), then discuss the
implications of two specific LSND analyses
\cite{Athanassopoulos:1997er,Auerbach:2003fz} for PNGBs that decay to
$e^+e^-$ pairs or $\mu^+\mu^-$ pairs, respectively.  The calculation
involves considerable uncertainties and assumptions, and should be
taken only as a sensitivity estimate --- a dedicated analysis by the
LSND collaboration is required to obtain a reliable limit.
However, this analysis demonstrates that a dedicated analysis by the
LSND collaboration would set some of the tightest constraints on PNGBs
that couple to both quarks and leptons.

We next discuss the production of dark gauge bosons in pion decay, and
the implications of the same two analyses for these models.  Our
results are consistent with \cite{Batell:2009di}, but we have
considerably elaborated the discussion of the experimental sensitivity.
The LSND sensitivity for this model overlaps closely with that of the
SLAC electron beam-dump experiment E137 \cite{Bjorken:1988as}, as
computed in \cite{Bjorken:2009mm}.

\subsection{Production of PNGBs at LSND and other Proton Beams}\label{ssec:protonBeam}

\begin{table}[t]
\begin{ruledtabular}
\begin{center}
\begin{tabular}{ccccccc}
 & {\small$E_{\rm beam}$ (GeV)} & $N_p$   &  $X_t$ (m)    &    $X_d$ (m)     &  $n_{\pi^0} \epsilon_{\rm geo}$  &   $\bar{E}_a$ (GeV) \\
\hline
\hline
CHARM \cite{Bergsma:1985qz} & 400& $2.4 \times 10^{18}$  & 480   & 515   &  0.12  & 25 \\
LSND \cite{Athanassopoulos:1996ds,Athanassopoulos:1997er,Auerbach:2003fz}  & $0.8 $ & $\sim 10^{23}$            & 29.7  &  38     &  see text			& 0.3 \\
MINOS / MINERvA \cite{Lebedev:2007zz,Minerva:2006zzb} & 120 & $3.8 \times 10^{20}$   & 1050 & 1087 &  0.0006  		& 20 \\
MiniBooNE \cite{AguilarArevalo:2008yp}  & 8.9 & $10^{21}$   & 541 & 553 & 0.002  	& 2.7 \\
\end{tabular}
\caption{Shown are the total number of incident protons $N_p$, the distance from the target to the open decay region in front of the detector (i.e.~the thickness of 
the shield) $X_t$, the distance from the target to the end of the detector $X_d$, the geometric 
acceptance times the number of pions per incident proton $n_{\pi^0} \epsilon_{\rm geo}$, and the 
median PNGB energy $E_a$.  These numbers were used to calculate the sensitivity of CHARM, 
MINOS/MINERvA  (we always use the larger MINOS detector for estimates), 
and MiniBooNE to PNGBs produced directly in the proton dump.
For LSND, we used a more involved procedure described in the text.
}
\label{tab:hadronic}
\end{center}
\end{ruledtabular}
\end{table}

In this section, we will consider the production and experimental sensitivity to pseudoscalars from proton beam dumps.  We will focus on the LSND experiment because it can set the most stringent limits, although the MINOS and MiniBooNE experiments can also set interesting limits. We will discuss them more extensively in \S \ref{sec:modern}.

Proton beam dumps can produce pseudoscalars directly through their mixing with pions.  If the pseudoscalars couple to quarks they will interact via the operator
\be
\frac{m_q}{F} a \bar q q \ \ \ \implies \ \ \  c \frac{m_\pi^2 F_\pi}{F} a \pi^0
\ee
where $c$ is an $O(1)$ parameter that depends on the up and down quark masses and any coefficients in the pseudoscalar coupling to the quarks.
Since the pseudoscalar mixes with the pion, for every pion that is produced through a QCD process there is a probability of approximately (for $c=1$) 
\be
\left( \frac{F_\pi}{F} \right)^2
\ee
of instead producing a PNGB.  

We can estimate the production rate within the detector acceptance and the PNGB momentum distribution by using the measured rates for $\pi^0$:
\be
N_a = \bigg( \frac{F_\pi}{F} \bigg)^2\, n_{\pi^0} \, N_p \, \epsilon_{\rm geo}.
\ee
Here, $n_{\pi^0}$ is the number of pions produced per incident proton, $N_p$ is the total number 
of protons, and $\epsilon_{\rm geo}$ is the geometric acceptance (the solid angle subtended by the detector at 
the target divided by the solid angle of the beam).  In practice, only the product $ n_{\pi^0} \, N_p \, \epsilon_{\rm geo}$ is relevant but we have tabulated estimates for $n_{\pi^0} \, \epsilon_{\rm geo}$ and $N_p$ in Table \ref{tab:hadronic} for various experiments, for the sake of comparison.
To determine the number of observable $e^+e^-$ pairs, we must also account for the probability of decaying in the detectable region (either inside the detector or in an open region upstream of the detector):
\be\label{eq:electronsfromPNGBs}
N_e = N_a \,\left( e^{-\f{X_t}{\gamma c\tau}} - e^{-\f{X_d}{\gamma c\tau}} \right) \simeq N_a \,\f{X_d-X_t}{\gamma c \tau} \, e^{-\f{L_0}{\gamma c\tau}},
\ee
where $X_t$ and $X_d$ are the minimum and maximum decay lengths (roughly, $X_t$ is the thickness 
of the shield and $X_d$ is the distance from the target to the end of the detector), 
and in the second expression we have assumed that $\gamma c \tau \gg X_d-X_t$.

\begin{figure*}[!t]
\begin{center}
\includegraphics[width=.7\textwidth]{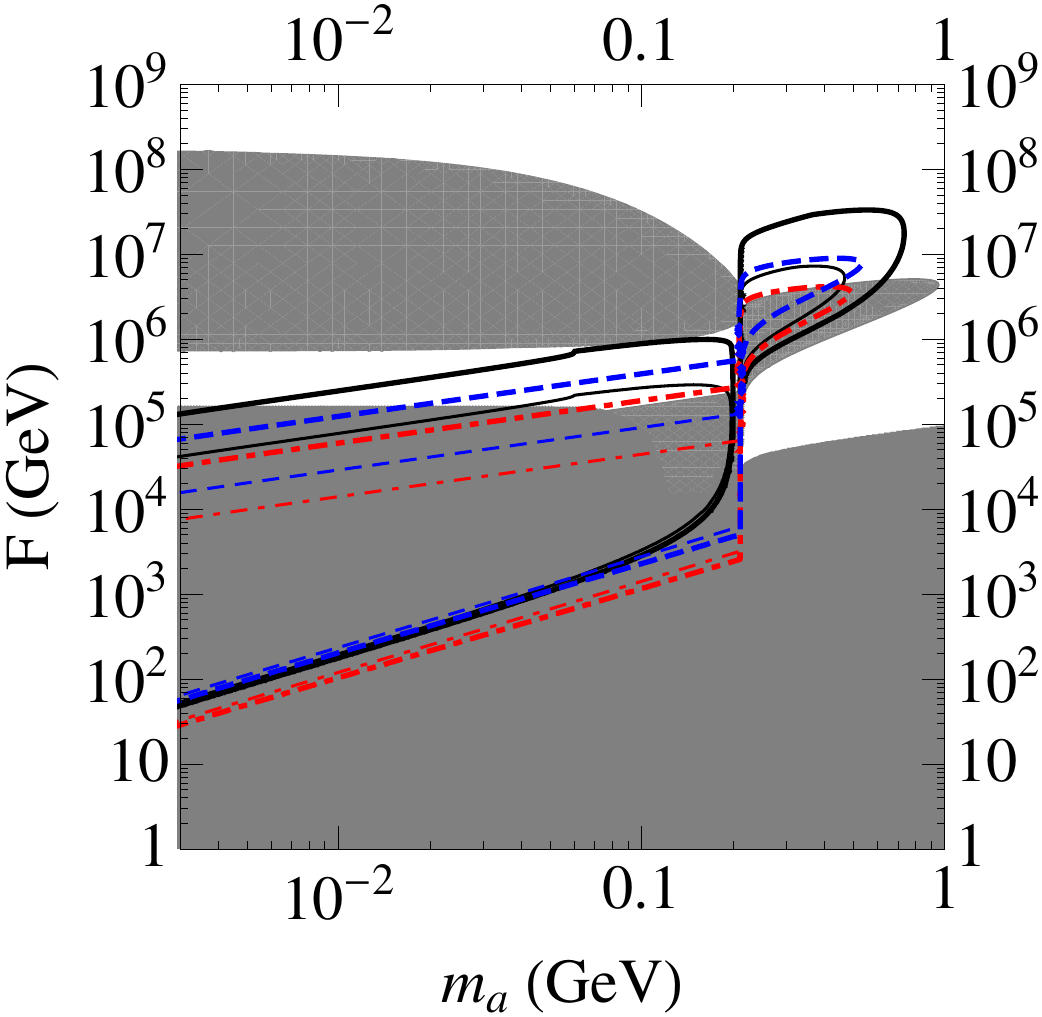}
\caption{Sensitivity of various neutrino experiments to pseudo-Nambu-Goldstone bosons 
as a function of their decay constants $F$ and their mass $m_a$.  
The thick (thin) black solid line corresponds to 10 (1000) events in LSND, the thick (thin) 
dashed blue line corresponds to 3 (1000) events in MiniBooNE, and the 
thick (thin) dot-dashed red line corresponds to 3 (1000) events in MINOS/MINERvA 
(in each case the inner regions correspond to more events than indicated by the line).
Here the PNGB is produced directly in the proton dump by a small mixing with the pion.  
For $m_a < 2 m_\mu$, the PNGB decays to an electron pair, while in the ``bubbles''  
for $m_a>2 m_\mu$ the PNGB decays predominantly to a muon pair. 
The gray shaded regions are the combined existing constraints from other beam dump 
experiments, meson decays, anomalous muon magnetic moment, and SN 1987a shown 
in the right plot of Fig.~\ref{fig:constraints}.  
}
\label{fig:comboHadroPlot}
\end{center}
\end{figure*}

\subsection{LSND Analyses Sensitive to PNGBs}
We now focus on two LSND results and their implications for PNGB's. 

\emph{\underline{PNGBs decaying to $e^+e^-$:}}
The analysis in \cite{Athanassopoulos:1997er} used $\sim 0.92 \times 10^{23}$ protons on target 
and looked for $\nu_\mu \to \nu_e$ oscillations using $\nu_\mu$ from $\pi^+$ decay in flight~\footnote{This analysis should not be confused with~\cite{Aguilar:2001ty} which looked for $\nu_\mu \to \nu_e$ oscillations from muon decays at rest and saw a the famous LSND excess. This analysis is not suitable for us because the search included a 2.2 MeV photon associated with the capture of the free neutron produced in the charged current reaction.}.  
The $\nu_e$ are detected via the inclusive charged-current reaction $\nu_e + ^{12}{\rm\! C} \to e^- + X$.  
This analysis focused on identifying electrons in the energy range 60 MeV to 200 MeV.  Various 
cuts were used in the analysis with an energy-dependent efficiency that is always near 10\%.  
Clearly this analysis should be sensitive to PNGBs decaying to electrons inside the LSND detector, 
although it is impossible to accurately estimate the efficiency without a dedicated analysis.  
One difference between our signal and the study in \cite{Athanassopoulos:1997er} is that in our case 
both an electron and a positron are produced in the detector, as opposed to just a 
single electron.  However, it is impossible to distinguish $e^+e^-$ events from a single 
electron (or single photon) event, and it has been suggested \cite{Bill} 
that we should assume that the total $e^+e^-$ pair energy 
(i.e.~the PNGB energy) would have been measured as the energy of a ``single-electron''.  
For simplicity, we will assume that the detection efficiency for an $e^+e^-$ pair is 
roughly the same as for the single electron analysis, i.e.~$\epsilon_{\rm eff,1} \sim 0.1$.

We have to estimate the energy distribution and number of PNGBs
incident on the LSND detector.  This should be roughly equivalent to
the number and energy distribution of pions, which has been simulated
with Monte Carlo by the LSND collaboration \cite{Bill}.  Specifically, we
will model the PNGB kinetic energy distribution using the predicted
pion kinetic energy distribution from \cite{Bill}, and rescaling the
total rate by $F_\pi^2/F^2$ from the total number of pions incident on the detector, 
$(8.6\pm 2.1)\times 10^{14}$ cm$^{-2}$.  This should be an excellent
approximation for $m_a \approx m_\pi$; it may be subject to additional
$\mc{O}(1)$ uncertainties for $m_a \ll m_\pi$ and especially for $m_a \gg
m_\pi$, which we neglect.
In addition to the reconstruction efficiency, we must also account for the fraction of PNGBs with
kinetic energy between $60-m_a$  and $200-m_a$, which for small $m_a$ and the assumed distribution 
is approximately $25\%$. 
We note that the mean of the pion kinetic energy distribution is at $\sim 275$ MeV with a root-mean-
square spread of $\sim 130$ MeV, so 
an analysis including higher-energy electrons would be significantly  more sensitive.

We show the 
number of $e^+e^-$ events obtained from PNGB decays in the $F$ versus $m_a$ parameter 
space in Fig.~\ref{fig:comboHadroPlot}, where the solid black thick and thin lines for 
$m_a < 2 m_\mu$ show 10 and 1000 $e^+e^-$ events, respectively.  
We have assumed that the number of PNGB decays inside the detector is given by 
Eq.~(\ref{eq:electronsfromPNGBs}) integrated over an PNGB energy from 
$60 {\rm~MeV} - m_a$ to $200 {\rm~MeV} - m_a$, and multiplied by the efficiency $\epsilon_{\rm eff,1}$.  
The analysis of \cite{Athanassopoulos:1997er} (see their Fig.~29) indicates that 
10 PNGB events in a 20 MeV energy bin below 200 MeV would have been easily noticed.  
We note that the sensitivity could have been increased by increasing the energy 
threshold for electrons beyond 200 MeV.

\emph{\underline{PNGBs decaying to $\mu^+\mu^-$:}}
We now turn our attention to the analysis in \cite{Auerbach:2003fz}, which would have been 
sensitive to $\mu^+\mu^-$ events.
The analysis in \cite{Auerbach:2003fz} considered $\sim 1.8\times 10^{23}$ protons on target 
and searched for $\pi^0 \to \nu_\mu \bar{\nu}_\mu$.  
The neutrinos interact with the nuclei in the detector to 
produce muons through the reactions $\nu_\mu + ^{12}{\rm\! C} \to \mu^- + p + X$ and  
$\bar{\nu}_\mu + ^{12}{\rm C} \to \mu^+ + n + X$.  This analysis focused on identifying high-energy,  
muon-like beam excess events in the energy range 160 MeV to 600 MeV electron equivalent.  
The muon was required to decay inside the detector.  Various selection cuts were used in the analysis 
\cite{Auerbach:2003fz} with an overall efficiency of $\sim 0.15$ for identifying 
neutrino interactions in the detector.  
This analysis should clearly have been sensitive to PNGBs decaying to $\mu^+\mu^-$ pairs 
inside the LSND detector, although it is impossible to accurately calculate the efficiency 
without a dedicated analysis.  
We simply estimate the efficiency for $\mu^+\mu^-$ pairs to be reconstructed as single-muon events to be $\epsilon_{\rm eff,2} \sim 0.1$, a little bit less than 
the efficiency for the original analysis. 
We estimate the PNGB kinetic energy distribution as described above. 
To obtain the total number of pions incident on the detector, we rescale the value we 
used above for the $a\to e^+e^-$ analysis to $1.7 \times 10^{15}$ cm$^{-2}$, since 
this analysis uses a factor of $\sim 2$ more protons.
Integrating over an PNGB energy from 
$160 {\rm~MeV} - m_a + 2 m_\mu$ to $600 {\rm~MeV} - m_a+2m_\mu$, we 
show the number of $\mu^+\mu^-$ events obtained from PNGB decays in the $F$ versus $m_a$ 
parameter space in Fig.~\ref{fig:comboHadroPlot}, where the solid black thick and thin 
lines for the region $m_a > 2 m_\mu$ show 10 and 1000 $\mu^+\mu^-$ events, respectively.
The analysis in \cite{Auerbach:2003fz} indicates that 10 PNGB events in a given energy bin 
would have been noticed, and so we believe that this is a reasonable number of events with 
which to set a tentative limit.   

\subsection{LSND Limits on Light Vector Bosons}

In \cite{Batell:2009di} it was shown that LSND may be sensitive to a new light vector boson $A'$ that mixes with $U(1)$ hypercharge; however, the precise sensitivity was unclear due to uncertainties in the LSND experiment.  We believe that we have clarified these uncertainties using information from \cite{Auerbach:2003fz,Athanassopoulos:1997er,Bill}.

With approximately $1.8 \times 10^{23}$ protons on target at $800$ MeV, the LSND experiment produced about $1.4 \times 10^{22}$ $\pi^0$'s.  The vast majority of these $\pi^0$'s decay to $\gamma \gamma$, so with a kinetic mixing parameter $\epsilon$ we expect a branching fraction of $2 \epsilon^2$ for $\pi^0 \to \gamma A'$.  We will only consider the simplest scenario, where the $A'$ 
decays to $e^+e^-$ with a lifetime
\be\label{eq:Alifetime}
\tau_{A'} = \frac{3}{\alpha \epsilon^2 m_{A'} \sqrt{1 - \frac{4 m_e^2}{m_{A'}^2}} \left(1 + \frac{2m_e^2}{m_{A'}^2} \right) }
\ee
that is also set by the kinetic mixing.
Due to the huge number of pions produced at LSND, the $A'$ may be visible even for very small $\epsilon$.  

To set a reliable limit, we use the results of the  $\pi^0 \to \nu \bar \nu$ search discussed previously~\cite{Auerbach:2003fz}.   They performed a Monte Carlo simulation to determine the flux of neutrinos in the LSND detector from this decay and plotted it in their Fig.~1 \cite{Auerbach:2003fz}.  In our case we want to consider $\pi^0 \to \gamma A'$, but in the limit that $m_{A'} \ll m_{\pi}^0$ the neutrino results should give a very good approximation to the distribution of $A'$ particles; beyond this limit we will continue to use the neutrino results with the caveat that we expect an $O(1)$ uncertainty in the rates.  Given this distribution of $A'$ particles, we computed the number of $A'$s that would have decayed inside the LSND detector using the lifetime formula above, and the $2 \epsilon^2$ branching ratio of $\pi^0 \to \gamma A'$.  

\begin{figure*}[!t]
\begin{center}
\includegraphics[width=.7\textwidth]{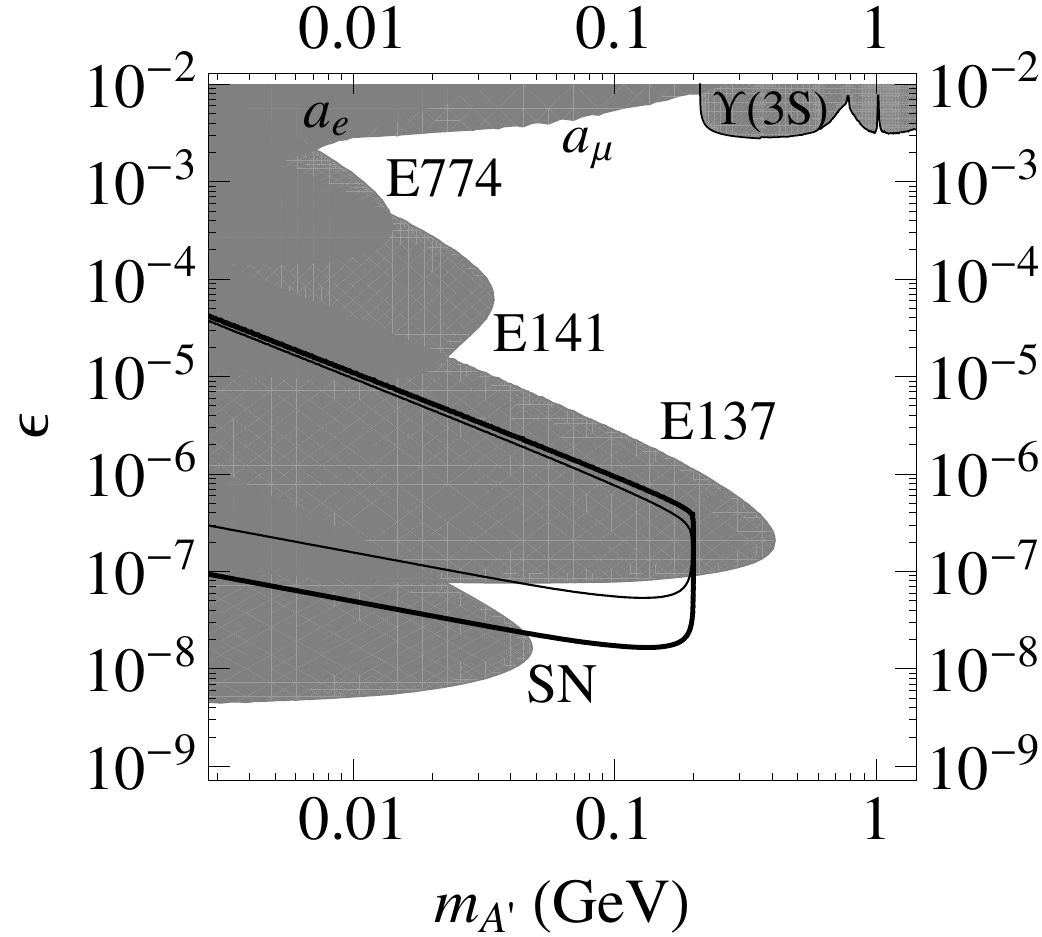}
\caption{The sensitivity of various beam dump, collider, and astrophysical probes of light vector bosons that kinetically mix with the standard model hypercharge, as a function of the kinetic mixing parameter $\epsilon$ and the vector boson mass $m_{A'}$ (from \cite{Bjorken:2009mm}).  
The thick (thin) solid black line corresponds to 10 (1000) signal events in LSND. 
The region enclosed by the thick line can be viewed as a very rough exclusion limit from the LSND experiment.  A re-analysis of the LSND data by the LSND collaboration could further extend the sensitivity of that experiment \cite{Batell:2009di}.  Further details are described in the text.  
}
\label{fig:vectorplot}
\end{center}
\end{figure*}

The only reliable data that can exclude an $A'$ decaying to $e^+ e^-$ inside the LSND detector is the analysis of \cite{Athanassopoulos:1997er}, which we discussed above in order to determine the LSND sensitivity to PNGBs.  From the study of $\pi^0$ backgrounds in \cite{Athanassopoulos:1997er} and from discussions with LSND collaborators \cite{Bill}, we expect that an $A'$ decay to $e^+e^-$ would have been interpreted as a single-electron event in the LSND detector.  The data on these types of events has only been published for energy depositions below $200$ MeV, and the analysis in question had an overall reconstruction efficiency of about  $10$\%.  Thus we use the Monte Carlo data \cite{Athanassopoulos:1997er, Bill} below this energy to set our tentative limits, which are displayed in Fig.~\ref{fig:vectorplot}.  Our limits are based on the conservative assumption that a signal of $50$ or more events after the $10$\% efficiency cut should have been visible.

\section{Sensitivity of Modern Neutrino Experiments}\label{sec:modern}

\subsection{Production from Proton-Nucleus Collisions}

We have discussed this production mode already in the context of LSND, in \S \ref{ssec:protonBeam}.  A similar analysis was also carried out by the CHARM experiment at CERN  \cite{Bergsma:1985qz}, a proton beam dump with a $400$ GeV beam, a $35$ m decay region, and a $3 \times 3$ meter detector.  Similar searches could be done using the MINOS or MINERvA detectors on the NuMI beamline, and with the MiniBooNE detector.  We summarize the parameters for these experiments in Table \ref{tab:hadronic}.  

In all cases we estimate the production of PGNBs by rescaling the pion production rate (determined either experimentally or from Monte Carlo) by a factor of $(F_\pi/F)^2$.  However, the production rate of very weakly coupled PGNBs could actually be significantly greater than this scaled rate because PGNBs produced within the target will virtually always escape from the target, whereas many of the pions produced by proton interactions might not make it out of the beam dump.  This issue may merit further exploration.

For the CHARM experiment, we take the experimental parameters from \cite{Bergsma:1985qz}. 
Our calculation of the PNGB exclusion region is shown in Fig.~\ref{fig:constraints} (right) and roughly 
agrees with  the results in \cite{Bergsma:1985qz}.

The MiniBooNE estimate is shown as a thick (thin) blue dashed line in Fig.~\ref{fig:comboHadroPlot}, 
corresponding to 3 (1000) events.  It is obtained as follows.  
We take the $\pi^0$ production cross section 
to be the average of the $\pi^+$ and $\pi^-$ production cross sections found in 
\cite{AguilarArevalo:2008yp}.  These cross sections are given as a function of angle and 
momentum, and one can show that 
the fraction of $\pi^0$'s produced in the target that point towards the 12 m detector 
at a distance of 541 m away is roughly $0.002$.  
In addition, \cite{AguilarArevalo:2008yp} gives the average number of 
$\pi^+$ and $\pi^-$ produced per ``particle-producing reaction'', which we take to be per 
incident proton on target -- we average these to find about 0.89 $\pi^0$'s per incident proton. 

The MINOS/MINERvA estimate is shown as a thick (thin) red dot-dashed line in 
Fig.~\ref{fig:comboHadroPlot}, again corresponding to 3 (1000) events.  
This estimate for MINOS was obtained by using data on their $\pi^+$ momentum distributions from 
Fig.~9.3 of \cite{Lebedev:2007zz}.  Based on the assumption that the momentum distribution for a 
general PGNB would be similar, we estimate from that figure that a fraction $0.0035$ of PGNBs 
produced by the NuMI proton dump would point towards the MINOS detector.  Data from 
\cite{Lebedev:2007zz} shows that there were roughly $0.18$ pions produced per proton 
on target, which allows us to estimate the total number of PGNBs produced.

Searches in either MiniBooNE and MINOS/MINERvA would have significant overlap with 
the existing CHARM result, perhaps slightly extending the region probed.  
However, the proposed ``Project X'' upgrades to NuMI, which would increase the beam intensity by a 
factor of $\sim 5$-$10$ and possibly include a larger near detector could extend sensitivity to larger $F$.

\subsection{Production through Muon $a$-sstrahlung}

\begin{table}[t]
\begin{ruledtabular}
\begin{center}
\begin{tabular}{cccccccc}
 & {\small$E_{\rm beam}$ (GeV)}  & {\small $N_\ell$}&  {\small $\bar{E}_\ell$ (GeV) }&  {\small$X_t$ (m)  }&{\small  $X_d$ (m)}  &{\small $W$ (m)} &  {\small$E_{\rm thresh}$ (GeV)} \\
\hline
\hline
E137  \cite{Bjorken:1988as} & 20.0  & $N_e = 1.2 \times 10^{20}$ & 20  &  180 & 380 & 3  & 2.0 \\
{\footnotesize{MINOS/MINERvA}} \cite{Anderson:1998zz}   & 120.0  & $N_\mu = 2.7 \times 10^{17}$ & 20  &  240 & 270 & 3  & 1.0 \\
MiniBooNE \cite{Stancu:2001dr}		&  8.9    & $N_\mu =2.0 \times 10^{17}$    & 1.3 &   450 & 462 & 12 & 0.1 \\
\end{tabular}
\caption{Shown are the proton beam energy $E_{\rm beam}$, the total number of incident 
electrons (for E137) or muons (for MINOS/MINERvA and MiniBooNE) $N_\ell$ and their average 
energy $\bar{E}_\ell$, the distance from the start of the electron or muon dump to the open decay 
region (if any) in front of the detector or to the detector itself (i.e.~the thickness of the shield) $X_t$, the 
distance from the start of the muon dump to the end of the detector $X_d$, 
the diameter of the detector $W$, 
and the threshold energy $E_{\rm thresh}$ to detect an electron or muon that originates from an 
PNGB decay. These numbers were used to calculate the sensitivity of E137 \cite{Bjorken:1988as}, 
MINOS/MINERvA \cite{Anderson:1998zz} (we always use the larger MINOS detector for 
estimates), and MiniBooNE \cite{Stancu:2001dr}
to PNGBs produced by bremsstrahlung off an electron beam (in the case of E137) or 
off a muon beam that is produced in a proton dump.  
}
\label{tab:lepto}
\end{center}
\end{ruledtabular}
\end{table}

A more unique search opportunity at MINOS/MINERvA and MiniBooNE
arises from their magnetic focusing of charged pions, which in turn
focuses the neutrinos and muons from their decay towards the
detectors.  As a result of this focusing, approximately one in a
thousand protons on target produces a muon that points toward the
neutrino detector.  As these muons stop in the rock upstream of the
detector, they can radiate very forward PNGB's in a
process that is analogous to ordinary bremsstrahlung.  
Unlike photon bremsstrahlung, however, 
the typical PNGB produced by this process carries a large 
fraction of the incident muon's energy.  
The production rate and kinematics can be
reliably calculated using the Weizs\"acker-Williams approximation
\cite{Tsai:1986tx}, where the nuclei in the fixed target provide an
effective photon beam.  We have relegated the details to Appendix \ref{app:muons},
so in this section we will simply give the results along with their
physical motivation.  

We will be interested in a wide range of values
for $m_a$; with this in mind we note that production is dominated by
emission angles \be \theta_a \lesssim \mathrm{max}
\left(\frac{m_a}{E_0}, \frac{m_\mu}{E_0} \right) \ee for $E_a \sim
E_0$.  This is very useful for estimating the angular acceptance,
which we compute by taking the ratio of the solid angle subtended
by the detector to the solid angle within $\theta_a$ of the beam
direction.  We also considered acceptance issues associated with the PNGB decay,
but because we require the PNGB to decay within the MiniBooNE
detector or in the relatively small open decay region in front of the MINOS/MINERvA detector, 
this is not important, and we ignore it.

\begin{figure*}[!t]
\begin{center}     
\includegraphics[width=.70\textwidth]{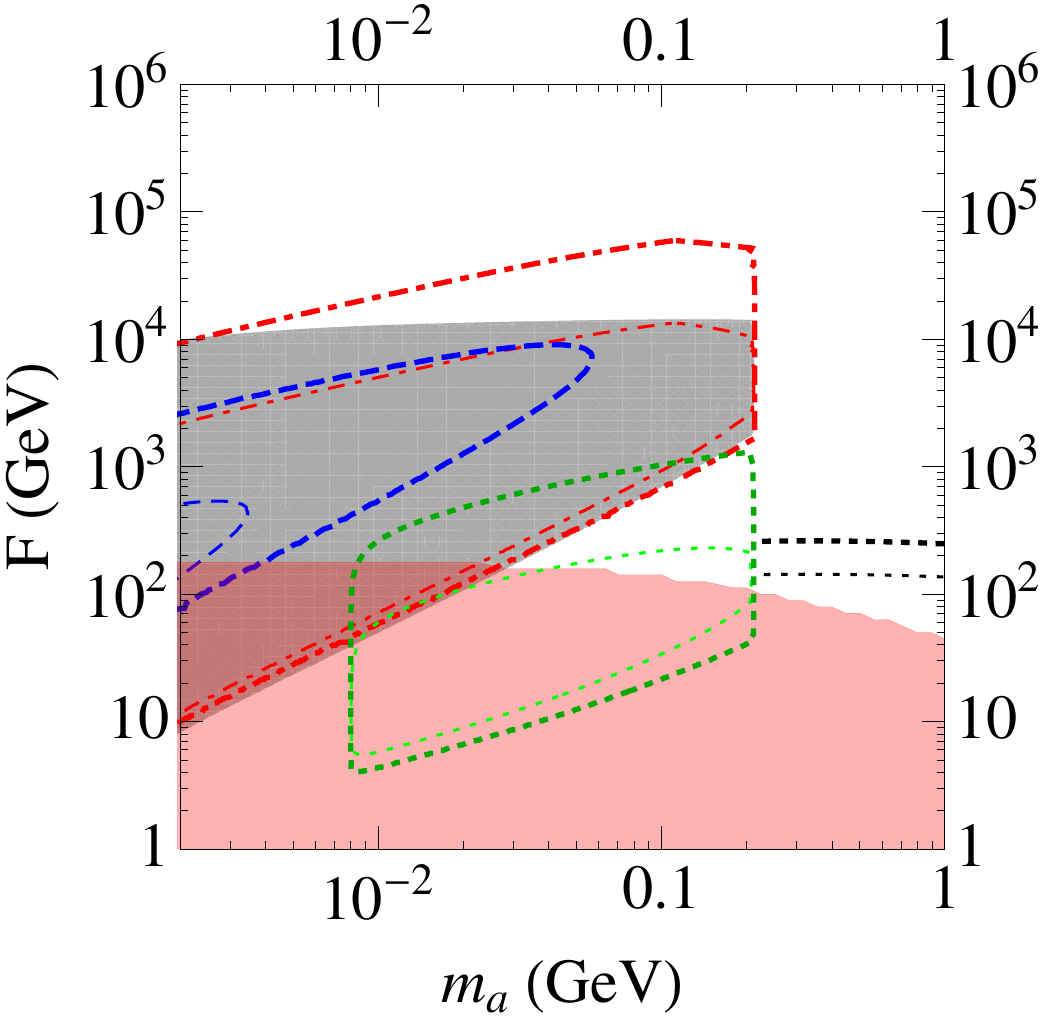}
\caption{
Sensitivity and constraints of various experiments to leptophilic 
pseudo-Nambu-Goldstone bosons as a function of their decay 
constants $F$ and their mass $m_a$.  
Here the PNGB is produced by bremsstrahlung off an incident muon 
or electron beam.  
Thick (thin) lines show rough sensitivity regions and 
correspond to 3 (1000) displaced $e^+e^-$ pairs in MINOS/MINERvA  
(red dot-dashed lines), MiniBooNE (blue dashed lines), and in a thin target experiment using the 
COMPASS muon beam (green dotted line, see \S \ref{sec:thin}).  
The thick (thin) dotted black lines correspond to $S/\sqrt{B} = 3~(10)$, where 
$S$ ($B$) are the number of prompt $\mu^+\mu^-$ signal (radiative background) events in COMPASS (see \S \ref{sec:thin}) (we have ignored the Bethe-Heitler background and the finite acceptance, 
so these lines should not be 
viewed as real significance lines but only as very rough estimates of what could be probed).
Inside the gray shaded region, E137 would have seen at least one event -- since they saw none, 
this region gives their approximate constraint.  
Details are described in the text.  
The light red region is the constraint from the muon anomalous magnetic 
moment (see \S \ref{sec:amu}).  
}
\label{fig:comboLepto}
\end{center}
\end{figure*}

The cross section is always peaked near $E_a \approx E_0$, even when $m_a \ll m_\mu$.  This is in stark contrast to the analogous formula for photon bremsstrahlung, where the rate is proportional to the inverse of the photon energy.  The difference is due to the contrasting soft emission behavior of gauge bosons and goldstone bosons -- the former have soft singularities, while emission of the latter vanishes in the soft limit due to their derivative couplings.  The peaking toward high energy fraction $E_a/E_0$ is further enhanced by a larger phase space when $m_a \gg m_\mu$.  The total cross section for pseudoscalar production from a muon beam has the parametric form
\be
\label{eq:mu-brem}
\sigma \approx \frac{m_\mu^2}{F^2} \frac{2 \alpha^2}{\mathrm{max}(m_\mu^2,m_a^2)}
\ee
Note that the formula in the case of an electron beam dump would be identical except with $m_\mu \to m_e$, so we see that muons are a much more efficient source of PNGBs as long as $m_a \gg m_e$.

To get a rough idea of our experimental reach, it is useful to have approximate formulas for the total production rates.  For a thin target, the yield is
\be
N_a \sim N_{\mu} \frac{m_\mu^2}{\alpha F^2 } \frac{T_e m_e^2}{\mathrm{max}(m_\mu^2, m_a^2)}
\ee
where $T_e$ is the number of electron radiation lengths of material.  In the case of a thick target, where the beam of muons is completely dumped, it is more difficult to give a simple parametric formula for the pseudoscalar yield because muons stop due to minimum ionization interactions, as opposed to bremsstrahlung.  However the equation above with $T_e \sim 100$ provides a rough estimate of the pseudoscalar yield for MINOS and MiniBooNE.

Fig.~\ref{fig:comboLepto} shows the number of $e^+e^-$ pairs in MINOS/MINERvA  
(red dot-dashed lines) and MiniBooNE (blue dashed lines)
as a function of the PNGB decay constant $F$ and the PNGB mass $m_a$.  
The thick (thin) lines correspond to 3 events (1000 events).
The gray shaded region corresponds to the approximate constraint from E137: inside the 
region, more than 1 $e^+e^-$ event would have been seen (note that we calculated this region 
using the procedure in \cite{Bjorken:2009mm}, but changing the couplings from an $A'$ to 
those relevant for PNGBs).  
Our estimate for the E137 constrained region agrees with that in \cite{Bjorken:1988as}.  
We see that MINOS/MINERvA can extend the E137 region, although the MiniBooNE region 
is contained within the E137 region since many PNGBs are produced in the dump with a large 
enough angle causing them to miss the detector (this is because 
the MiniBooNE muon beam has a lower energy 
and their detector is further away compared to MINOS/MINERvA). 
Note that these lines have been calculated with the full formula as detailed 
in the appendix.
The NOVA experiment and ``Project X" upgrades to the NuMI beamline 
will have a factor of 5--10 more protons and so will significantly extend the 
reach.  
The light red region is the constraint from the muon anomalous magnetic 
moment, while the other lines will be discussed in \S \ref{sec:thin}.

\section{Muon Fixed-Target Experiments with Thin Targets -- COMPASS}\label{sec:thin}

In this note, we have argued that new light states may be produced by bremsstrahlung off a dumped muon beam. Here we will briefly comment on the potential of using a fixed target setup to constrain weakly coupled light states. Compared to a dump experiment, the target will be much thinner, and we can thus search for particles with a much shorter lifetime. 

Electron fixed target experiments have been used to probe new light states~\cite{Bjorken:2009mm,Reece:2009un,Freytsis:2009bh,Essig:2010xa}, but muon fixed-target setups are different in several respects. The intensity of muon beams is obviously lower, but this may be partly compensated for by a much thicker target --- a muon beam can easily traverse a meter of material leaving a relatively quiet off-beam environment. Furthermore, muon beams will have an obvious advantage over electrons in PNGB searches, as their couplings are proportional to particle mass.  For example, it is instructive to consider the COMPASS experiment at CERN \cite{Abbon:2007pq}, where a 160 GeV muon beam strikes a low-$Z$ polarized target (they have also used higher-$Z$, but much thinner, targets). 
A detailed description of their polarized target is given in \cite{Abbon:2007pq}, and 
we approximate it as a 130 cm long target consisting of Lithium, with a packing factor of 0.5 (i.e. a column density of 34.7 g/cm$^2$, or 0.42 radiation lengths).  
In total, we estimate that this experiment has collided about $\sim 10^{15}$ muons.
Here we will illustrate the rough reach of this COMPASS-like setup assuming the Lithium target.  We have 
checked that a higher $Z$ target, such as Tungsten, of a similar thickness  would probe 
a somewhat larger parameter space, including high-$F$ regions above the muon threshold.  
We will not show this in any figures, focusing instead on the data set already  
collected by COMPASS. 

There are two regions of parameter space to consider, one in which $F$ is low and the PNGB decays 
promptly, and one in which $F$ is high and the PNGB decays with a displaced vertex. In the case of a 
prompt decay, one must search for a peak in the di-muon invariant mass on top of a sizable standard 
model background (decays to electrons may not be searched for this way because their interactions  in the target considerably degrade the mass resolution).
The signal to background ratio (where the background comes from di-lepton production via an 
off-shell photon) for such a search is roughly~\cite{Bjorken:2009mm}
\begin{equation}\label{eq:SoverB}
\frac{S}{B}\sim \frac{m_\mu^2}{F^2}\frac{3}{4\alpha^2}\frac{m_a}{\delta m}\,,
\end{equation}
where $\delta m$ is the resolution-limited mass window used in the search.  Based on the angular and momentum resolutions for COMPASS~\cite{Ageev:2005ud}, we infer a mass resolution $\sigma_m = 11$ MeV, making a window width $\delta m \approx 2.5 \sigma_m = 27$ MeV appropriate for estimating sensitivity.
Additional two-photon diagrams (which we refer to as the ``Bethe-Heitler'' background) and their interference have been neglected; these actually dominate over the ``radiative'' dilepton production assumed in \ref{eq:SoverB}, but can be reduced to an order-1 fraction of the background by kinematic selection.  

We can combine Eq.~(\ref{eq:SoverB}) with the expected number of signal events using the formulas in 
Appendix \ref{app:muons} to get a very rough estimate for the sensitivity of such a search.  These estimates omit several factors, all of which degrade sensitivity: the finite detector acceptance, the additional Bethe-Heitler background, and the finite acceptafinite of kinematic selection needed to veto the dominant Bethe-Heitler background.  Assuming $10^{15}$ muons on target 
and a mass window $\delta m = 27$ MeV~\cite{Ageev:2005ud}, we show the 
``3$\sigma$'' and ``$10\sigma$'' estimated reach in Fig.~\ref{fig:comboLepto} (the neglected factors are expected to reduce the sensitivity each point in parameter space by a factor of 2--4). 
Such a search in existing COMPASS data (or that of a similar experiment) would be sensitive to leptophilic PNGBs with $F$  near the weak scale and masses between the dimuon threshold and a few GeV.  
Higher $F$'s can be probed by increasing the target thickness in radiation lengths, for example by using a high-$Z$ target of comparable length.

Another parameter range that can be studied by muon fixed-target experiments is that where the PNGB decays are significantly displaced. If the vertex can be 
reconstructed to be downstream of the target region, beyond the tails of Standard Model backgrounds, a few events could be enough to claim a discovery.  
The transverse vertex resolution of COMPASS is about 0.1 mm \cite{Abbon:2007pq}, so that 
the vertex resolution in the 
direction of the beam is about $\sigma_z = (0.1 ~{\rm mm}) \times E_{\rm beam}/m_a$.  
In Fig.~\ref{fig:comboLepto}, we show the area of parameter space that yields 3 or 1000 events in the 
region between $5\sigma_z$ and 10 meters behind the 130 cm Lithium target 
(for this estimate we assume that the PNGB is produced in the middle of the target). 
Note that the vertex resolution is worse for lower-mass PNGBs, so that for 
$m_a \lesssim 10 MeV$, we have $5\sigma_z \gtrsim 10$ m, and there is no good search region 
(which is why the plot cuts off for low $m_a$).
An analysis of the data in COMPASS, or a similar experiment of this type, can cover new ground with 
respect to E137 or the neutrino experiments at lower $F$, and can close the window between the latter 
searches and the muon $g-2$ limit. With a long higher-$Z$ target such as Tungsten, more PNGBs would be produced so that some region of 
parameter space at $F\sim 10^4$ GeV above the muon threshold may also give rise to observably large rates of displaced decays.

In summary, the very rough estimates above suggest that  a fixed target experiment with a focused muon beam may be able to probe unconstrained regions in PNGB parameter space. 
In fact, the existing data set of the COMPASS experiment may already be able to set some 
interesting new limits for leptophilic PNGBs.  
It may also be worthwhile to consider using more diffuse muon ``beams'' such as those in neutrino factories in a thin target setup due to their higher intensity.  We leave this for future thought.

\section{Conclusions}\label{sec:conclusions}

We have explored the sensitivity of neutrino oscillation experiments to three types of new light states -- 
vector bosons that kinetically mix with the photon, pseudoscalars that couple to quarks and leptons, and 
pseudoscalars that couple preferentially to leptons.  The first two are strongly constrained by rare 
decays, fixed target experiments, and supernova 1987a (all of which we have reviewed), 
whereas there are fewer tests of the third class.  

The sensitivity of the LSND experiment to vector bosons was discussed in \cite{Batell:2009di}, but the details of the LSND detector and analyses were not considered.  We have shown that the analyses of \cite{Auerbach:2003fz, Athanassopoulos:1997er} would have been sensitive to vector bosons with mass below $2 m_\mu$ and a large range of coupling strengths.  This sensitivity has significant overlap with the E137 experiment \cite{Bjorken:1988as}, but LSND does probe a new region at very weak coupling, and in any case the LSND results serve as an important cross-check.  Because LSND dumped a larger number of protons, other neutrino experiments are not as sensitive to the light vector boson scenario.

We also considered the sensitivity of various neutrino experiments to pseudo-scalars that couple to quarks, so that they can be produced in proton beam dumps.  LSND remains the most sensitive experiment for pseudoscalars with mass below $2 m_\mu$, nearly closing the gap between fixed target experiments and supernova constraints.  However, other experiments are more sensitive to heavier pseudoscalars.  The MiniBooNE and MINOS experiments are currently competitive with the best limits on these particles.  However, the estimates that we derive for these experiments may be too conservative because we estimate the rate of PGNB production by scaling the pion production rate, which may be an underestimate if many pions get stuck in the beam dump.  This issue merits further study.

The MINOS and MiniBooNE experiments produce neutrinos from focused muon beams; the requisite muon beam dumps provide a unique opportunity to search for pseudoscalars that couple preferentially to leptons, since we expect these particles to couple far more to muons than to electrons.  The MINOS experiment is sensitive to leptophilic pseudoscalars with decay constants almost an order of magnitude greater than any previous experiment (in particular E137), while MiniBooNE can probe a region that is contained within that of E137. 

A thin target experiment with a muon beam, such as that available in COMPASS, offers a unique 
probe for leptophilic PNGBs.  An analysis using the existing COMPASS data set and 
looking for either $e^+e^-$ pairs originating from displaced vertices behind their (Lithium) target 
or for a spike in the $\mu^+\mu^-$ invariant mass spectrum of muons coming from the target 
should be sensitive to new regions of parameter space.  A similar experiment using a higher-$Z$ 
target would have even more sensitivity. 

Upgrades to the NuMi beamline followed by the proposed ``Project X'' experiment will explore new 
parameter space for both standard and leptophilic pseudoscalars.  It is our hope that in the future these 
experiments will perform dedicated analyses to explore and constrain new weakly coupled low-mass 
particles.

\section*{Acknowledgements}
We thank Philip Schuster for collaboration in early stages of this work.  We also thank 
James Bjorken, Joe Lykken, Chris Polly, Geoff Mills, Simona Murgia, Michael Peskin, Ronald Ransome, 
Brian Rebel, Byron Roe, David Schmitz, Tomer Volansky, Jay Wacker, Hywel White, and 
Geralyn Zeller  for very useful discussions.
We especially want to thank William Louie for extensive discussions and correspondence 
about the LSND detector and publications.
RE and JK are supported by the US DOE under contract number DE-AC02-76SF00515. Fermilab is operated by Fermi Research Alliance, LLC, under Contract DE-AC02-07CH11359 with the United States Department of Energy. 
We acknowledge the hospitality of the Aspen Center for Physics where part of this work was 
done.

\appendix

\section{Pseudoscalar Production}\label{app:muons}

We are interested in the pseudoscalar production rate from muons braking in a fixed target.  This 
process is analogous to ordinary bremsstrahlung, and it can be reliably calculated using the 
Weizs\"acker-Williams approximation \cite{Tsai:1986tx}, where the nuclei in the fixed target provide an 
effective photon beam.  When the incoming muon has energy $E_0$, the differential cross section to 
produce a pseudoscalar of mass $m_a$ with energy $E_a = x E_0$ is
\be\label{eq:cs}
\frac{d \sigma}{dx~d\!\cos\theta_a} = \frac{m_\mu^2}{F^2} \frac{2 \alpha^2 E_0 x}{U^2} \left[x^2 - \frac{2 m_a^2 x(1-x)}{U} + \frac{2 m_a^2}{U}(m_a^2 (1-x)^2 + m_\mu^2 x^2 (1-x)) \right] \,\chi
\ee
where 
\be
U = E_0^2 \theta_a^2 x + m_\mu^2 x + m_a^2 \frac{1-x}{x}
\ee
is the virtuality of the intermediate muon in initial state bremsstrahlung, and $\chi$ is a form factor 
that can be found in \cite{Tsai:1986tx}.
We will be interested in a wide range of values for $m_a$; with this in mind we see that production is dominated by
\be
\theta_a \lesssim \mathrm{max} \left(\frac{m_a}{E_0}, \frac{m_\mu}{E_0} \right)
\ee
for $x \sim 1$.  For angles larger than this, the differential cross section falls off rapidly, as $1/\theta_a^4$, so this angular scale sets the width of the PNGB beam for the purposes of angular acceptance.  We compute the angular acceptance by taking the ratio of the solid angle subtended
by the detector to the solid angle within $\theta_a$ of the beam
direction. 

Integrating Eq.~(\ref{eq:cs}) over the angle $\theta_a$, we find
\be
\frac{d \sigma}{dx} = \frac{m_\mu^2}{F^2} \frac{2 \alpha^2}{m_\mu^2} x \left[ \frac{ 1 + \frac{2}{3}f}{(1+f)^2} \chi_1(Z)
+ \left( \frac{1}{3 f^2}(1+f) \log(1+f) - \frac{1+4f+2f^2}{3f(1+f)^2} \right) \chi_2(Z) \right]
\ee
where
\bea
f & = & \frac{m_a^2(1-x)}{m_\mu^2 x^2}
\eea 
and we have now included the form factors
\bea
\chi_1(Z) & = & Z^2 \ln(184\, Z^{-1/3}) + Z \ln (1194\, Z^{-2/3}) \\
\chi_2(Z) & = & Z^2 + Z.
\eea
It is important to note that the cross section is always peaked near $x \sim 1$, even when $m_a \ll m_\mu$.  This is in stark contrast to the analogous formula for bremsstrahlung, which is proportional to $1/x$.  The difference is due to the different soft emission behavior of gauge bosons and goldstone bosons -- the former have soft singularities, while emission of the latter vanishes in the soft limit.  The cross section is dominated for $x \sim 1$ with the parametric form
\be
\sigma \approx \frac{m_\mu^2}{F^2} \frac{2 \alpha^2}{\mathrm{max}(m_\mu^2,m_a^2)}
\ee
The max$(m_\mu^2, m_a^2)$ factor comes from the presence of the function $f \sim m_a^2/m_\mu^2$ in the denominator for larger $m_a/m_\mu$.
The formula in the case of electrons would be identical with $m_\mu \to m_e$, so we see that as claimed, muons are a much more efficient source of PNGBs as long as $m_a \gg m_e$.

To use these formulae we must account for the way that the muon slows in a beam dump.  The number of PNGBs produced per incident muon is
\be\label{eq:yield}
\frac{dY}{dx} = \frac{N_A X_e}{A} \int_{E_a}^{E_0} d E_1 \int_0^{T_e} dt_e I_{\mu} (E_0, E_1, t_e) \frac{d \sigma}{dx'}
\ee
where $E_0$ is the energy of the original incident muons, $E_a$ is the energy of the produced PNGBs, $x' = E_a / E_1$, $x = E_a / E_0$, $N_A$ is Avogadro's number, $X_e$ is the unit (electron) radiation length in g/cm$^2$,  $A$ is the atomic number of the material, and $T_e$ is the total number of (electron) radiation lengths in the target or beam dump.  The function $I_\mu (E_0, E_1, t_e)$ is the distribution of muon energies after the muons have traversed $t_e$ radiation lengths.  

It is essential to remember that muons stop primarily through minimum ionizing interactions, and not through radiation, so the number of radiation lengths is not directly related to the muon energy loss.  We therefore use electron radiation lengths in our computations, so that $t_e$ and $T_e$ can be much greater than $1$ without completely depleting the energy of the muon beam.  We estimate the function $I_\mu$ using the relevant material properties (if the intervening material is earth, we use Silicon).

To compute the number of electron or muon pairs from PNGB decays in a detector, 
we simply integrate Eq.~(\ref{eq:yield}) 
times the probability that an PNGB with this energy decays in (or in some cases in front of) the relevant 
detector (see Eq.~(\ref{eq:electronsfromPNGBs})).

\bibliography{paperv1}
\end{document}